\documentstyle[10pt,aaspp4,epsf]{article} 
 
\def\psim{\lower.5ex\hbox{$\; \buildrel \propto \over\sim \;$}} 
\def\gtrsim{\lower.5ex\hbox{$\; \buildrel > \over\sim \;$}} 
\def\lesssim{\lower.5ex\hbox{$\; \buildrel < \over\sim \;$}} 
\def\gm{\gamma_m} 
\def\g2{\gamma_2} 
\def\gc{\gamma_c} 
\def\elnumax{L_{\nu,{\rm max}}} 
\def\e{{\epsilon}}
\begin{document} 
 
\title{Spectral Energy Distributions of Gamma Ray Bursts\\ Energized by External Shocks} 
 
\author{Charles D. Dermer\altaffilmark{1}, Markus 
B\"ottcher\altaffilmark{2,3}, \&  James Chiang\altaffilmark{4}} 
 
\altaffiltext{1}{E. O. Hulburt Center for Space 
Research, Code 7653, 
       Naval Research Laboratory, Washington, DC 
20375-5352} 
\altaffiltext{2}{Department of Space Physics and 
Astronomy,  Rice University, Houston, TX  77005-1892} 
\altaffiltext{3}{Chandra Fellow} 
\altaffiltext{4}{JILA, University of Colorado, 
Boulder, CO 80309-0440}

\begin{abstract}  
 
Sari, Piran, and Narayan have derived analytic formulas to model the spectra from gamma-ray burst blast waves that are energized by sweeping up material from the surrounding medium.  We extend these expressions to apply to general radiative regimes and to include the effects of synchrotron self-absorption.  Electron energy losses due to the synchrotron self-Compton process are also treated in a very approximate way. The calculated spectra are compared with detailed numerical simulation results. We find that the spectral and temporal breaks from the detailed numerical simulation are much smoother than the analytic formulas imply, and that the discrepancies between the analytic and numerical results are greatest near the breaks and endpoints of the synchrotron spectra.  The expressions are most accurate (within a factor of $\sim 3$) in the optical/X-ray regime during the afterglow phase, and are more accurate when $\epsilon_e$, the fraction of swept-up particle energy that is transferred to the electrons, is $\lesssim 0.1$. The analytic results provide at best order-of-magnitude accuracy in the self-absorbed radio/infrared regime, and give poor fits to the self-Compton spectra due to complications from Klein-Nishina effects and photon-photon opacity.  \end{abstract} 
 
\section{Introduction} 
 
The discovery of X-ray and long wavelength afterglow radiation from gamma ray bursts (GRBs) finds a convincing explanation through the blast-wave model (e.g. Paczy\'nski \& Rhoads \markcite{pr93}1993; M\'esz\'aros \& Rees \markcite{mr97}1997). In this model, a coalescence or collapse event, possibly involving the formation of a black hole, produces a weakly baryon-loaded fireball with directional energy releases $\partial E/\partial\Omega$ as large as $\sim 10^{53}$ ergs sr$^{-1}$.  The blast wave sweeps up material from the surrounding medium during its expansion.  The swept-up particles are a source of free energy in the comoving blast wave frame.  Depending on the magnetic field strength and the intensity of the surrounding radiation field, this energy is primarily radiated through synchrotron and Compton processes. Particle escape, or other channels of particle energy loss such as adiabatic losses or secondary production for the high-energy protons (Pohl \& Schlickeiser \markcite{ps00}2000), may dominate at certain times in the blast wave evolution and in certain regimes of particle energy. 
 
Sari, Piran, \& Narayan (\markcite{spn98}1998; hereafter SPN98) have derived useful formulas to model the emissions from a decelerating blast wave in the afterglow phase, and asymptotic forms were given in the adiabatic and radiative limits of blast-wave evolution.  These formulas have been widely used to interpret observations of GRB afterglows. Their work was restricted to the optical and X-ray range where synchrotron self-absorption (SSA) is generally unimportant.  Moreover, the synchrotron self-Compton (SSC) process was not treated, and no comparison of the analytic expressions with detailed numerical simulations was carried out to assess, for example, the inaccuracy incurred by not considering adiabatic losses on the evolution of the electron energy spectra. 
 
In this article we generalize the formulas of \markcite{spn98}SPN98 to include SSA and SSC processes and to apply to arbitrary radiative regimes in the evolution of the the GRB blast wave. The results are compared with a detailed numerical model (Chiang \& Dermer \markcite{cd99}1999; Dermer, Chiang, \& Mitman \markcite{dcm99}1999).  The numerical treatment approximates the blast wave as a thin shell within which the energized particles radiate in a uniform magnetic and photon field.  It therefore applies to the forward shock, which is the appropriate system that should be compared with the analytical formulas. Emissions from the reverse shock are not considered here. 
 
In the next section, we summarize the parameters that enter into the standard relativistic blast-wave model for GRBs, and describe the electron distribution function and evolution of the blast wave.  In Section 3, the SSA process is considered.  Section 4 describes our treatment of SSC processes.  A presentation of the results of the semi-analytic model, including a comparison with numerical data, is provided in Section 5.  We summarize in Section 6.  
 
\section{Electron Distribution and Blast Wave Evolution} 
 
We adopt the notation of \markcite{spn98}SPN98, except where indicated.  The power injected in nonthermal electron energy in the comoving blast-wave frame is assumed to be equal to a constant fraction $\e_e$ of the power $\dot E$ of particles swept-up by the blast-wave.  For an uncollimated blast wave expanding isotropically into a uniform medium with density $n$, $\dot E = 4\pi r^2  n m_p c^3 \beta (\Gamma^2 - \Gamma)$ (Blandford \& McKee \markcite{bm76}1976).  Here $\Gamma$ is the blast-wave Lorentz factor and $c \beta = c(1-1/\Gamma^2)^{1/2}$ is its speed.  The electrons are assumed to be injected in the form of a power law with index $p$ and with minimum and maximum Lorentz factors $\gm$ and $\gamma_2 \cong \e_{\rm max}\cdot 4\times 10^7/[B{\rm (G)}]^{1/2}$, where the latter expression derives from a comparison between the maximum electron acceleration rate and the synchrotron loss rate, and $\e_{\rm max}\leq 1$. We set $\e_{\rm max} = 1$ in the examples shown in this paper.  If the number of energized nonthermal electrons is equal to the number of swept-up electrons, then 
\begin{equation} 
{p - 1 \over p-2 } \; \bigl({\gm^{2-p}-\g2^{2-p}\over \gm^{1-p}-\g2^{1-p}}\bigr) = 1 + \e_e(\Gamma - 1)m_p/m_e\; , 
\label{gmin} 
\end{equation} 
which reduces to the commonly used expression $\gm \cong \e_e [(p-2)/(p-1)](\Gamma - 1)m_p/m_e$ when $\g2\gg\gm\gg 1$ and $p> 2$.  Eq.\ (\ref{gmin}) is easily solved numerically to give an accurate value for $\gm$. Note that the meaning of $\epsilon_e$ in the numerical simulation differs from its meaning in the analytic model (Wijers \& Galama \markcite{wg99}1999); in the latter case, $\epsilon_e$ represents the ratio of (residual) nonthermal energy in electrons to that in protons.  
 
Following the reasoning of \markcite{spn98}SPN98, electrons will cool in the comoving fluid frame by synchrotron and Compton losses to Lorentz factor  
\begin{equation} 
\gc = {3m_e \over 4\lambda m_p n c\e_B  \sigma_t \Gamma^3 t (1+f_{\rm KN}u_{\rm ph}/u_B)}\; 
\label{gcool}  
\end{equation} 
by observing time $t$. Here $\lambda$ is the compression ratio (set equal to 4 in the following calculations), and $u_B = B^2/8\pi = \lambda n m_pc^2 \e_B \Gamma^2$ and $u_{\rm ph}$ are the magnetic field and photon energy densities, respectively.  When Klein-Nishina effects on Compton scattering are important, $u_{\rm ph}$ is corrected for the reduction in the scattering efficiency using the factor $f_{\rm KN}$ ($\leq 1$), as discussed in more detail below. 
 
The distribution of energized electrons is approximated by 
\begin{equation} 
N(\gamma ) \cong (s-1) N_e \gamma_0^{s-1} 
\, \cases{ \gamma^{-s}\;, & for $\gamma_0\leq\gamma\leq\gamma_1$ \cr 
	 \gamma_1^{-s}(\gamma/\gamma_1)^{-(p+1)} & for $\gamma_1 \leq \gamma \leq \gamma_2$,\cr} 
\label{edist} 
\end{equation} 
where $N_e = 4\pi r^3 n/3 $ is the total number of nonthermal electrons.  In the slow cooling limit, $\gamma_0 = \gm$, $\gamma_1 = \gc$, and $s=p$, whereas in the fast cooling limit $\gamma_0 = \gc$, $\gamma_1 = \gm$, and $s = 2$.  The blast wave's radiative efficiency $\e = \e_e\e_{\rm rad}$, where $\e_{\rm rad} \cong 1$ in the fast cooling limit, and $\epsilon_{\rm rad} = (\gm/\gc)^{p-2}$ in the slow cooling limit (Moderski, Sikora, \& Bulik \markcite{msb99}1999).  
 
The evolution of $\Gamma$ in radiative regimes bridging the adiabatic and radiative limits can be quickly obtained by numerically solving the equation of motion $d\Gamma/dm = -(\Gamma^2 -1)/M$. Here $dm = 4\pi r^2 m_p n dr$ and $M = \int dm\; [\e +\Gamma(1-\e)]$ is the total energy in the blast wave, including both rest mass and kinetic energy. This prescription for the change in $M$ is consistent with energy conservation at all energies, and reduces in the nonrelativistic limit to Oort's solution based on momentum conservation rather than the Sedov self-similar solution, as noted by Huang, Dai, \& Lu ({\markcite{hdl99}1999).  Analytic solutions for these equations when $\e$ is constant are presented by B\"ottcher \& Dermer (\markcite{bd00}2000), where observational implications of general radiative regimes are discussed. 
 
\section{Synchrotron Self-Absorption} 
 
In the absence of SSA, the analytic approximation to the synchrotron radiation consists of a low-energy hard power-law emission spectrum with spectral luminosity $L_\nu \propto \nu^{1/3}$ at $\nu < \nu_0$, connected to an intermediate branch with $L_\nu \propto \nu^{-(s-1)/2}$ at $\nu_0 < \nu < \nu_1$, which joins to a higher energy branch with $L_\nu \propto \nu^{-p/2}$ at $\nu_1 < \nu < \nu_2$.  The  frequencies of the breaks in the synchrotron spectrum are given by $\nu_i = \Gamma\gamma_i^2 eB/(2\pi m_e c)$, with $i=0,1,2$. (Observed frequencies should be divided by the factor $(1+z)$ to implement a redshift correction for a sourceat redshift $z$.) When SSA is important, the spectrum exhibits a break at $\nu < \nu_a$, where $\nu_a$ is the self-absorption frequency.  If $ \nu_a < \nu_0$, then $F_\nu \propto \nu^2$ at $\nu \ll \nu_a$.  When $\nu_a > \nu_0$,  $F_\nu \propto \nu^2$ at frequencies $\nu < \nu_0$ and $F_\nu \propto \nu^{5/2}$ for $\nu_0 < \nu < \nu_a$. We do not treat the case $\nu_a > \nu_1$, though it is a straightforward generalization of the results presented here. 
 
When $\nu < \nu_0$, we calculate the SSA coefficient from the standard expression 
\begin{equation} 
\kappa_{\nu^{\prime}}= -{1\over 8\pi m_e \nu^{\prime 2}V_{\rm bw}}\int _1^\infty d\gamma P(\nu^\prime,\gamma)\gamma^2 {\partial\over \partial\gamma} [{N(\gamma)\over \gamma^2}]\;, 
\label{SSAcoef} 
\end{equation} 
where $N(\gamma)\cong (s-1)N_e \gamma_0^{s-1}\gamma^{-s}$ from eq. (\ref{edist}), the blast-wave volume $V_{\rm bw} = 4\pi r^2 \Delta^\prime$, and $\Delta^\prime$ is the comoving-frame width of the shell. In the specified limit, the calculation of the absorption coefficient is simplified because we can use the  asymptotic form for the synchrotron emissivity function  
\begin{equation} 
P(\nu ) = {3^{1/2} e^3 B\sin \alpha \over m_e c^2} F(\nu/\nu_{cr}) \rightarrow {4\pi e^3 B\over \Gamma(1/3) m_e c^2}({\nu \over 3\nu_B\gamma^2})^{1/3}\; , 
\label{sync} 
\end{equation} 
where the right-hand-side applies in the limit $\nu\ll \nu_{cr}$, where $\nu_{cr} = 3\nu_B\gamma^2/2$ is the critical frequency. The exact expression for $F(x)$ is given by, e.g., Rybicki \& Lightman (\markcite{rl79}1979). Note that $\Gamma(1/3) \cong 2.679$ is a Gamma function. In this expression $\nu_B \equiv eB\sin\alpha /2\pi m_ec$ and we let the pitch angle $\alpha = \pi/2$. Substituting eq.({\ref{sync}) into eq.(\ref{SSAcoef})and letting the relation $\kappa_{\nu^{\prime}}\Delta^\prime =1$ define the comoving-frame absorption frequency, we obtain 
for the observed SSA frequency the expression 
\begin{equation} 
\nu_a ({\rm Hz}) = 9.9\; [{(s+2) (s-1)\over (s+2/3)}]^{3/5} \; (rn)^{3/5} B^{2/5}\;{\Gamma\over\gamma_0},\; {\rm when} \; \nu_a < \nu_0\; . 
\label{nuabs} 
\end{equation} 
All of the terms in eq.(\ref{nuabs}) are in cgs units.
 
For the case $\nu_0 < \nu_a < \nu_1$, we use the results quoted by Gould (\markcite{gould79}1979) for the absorption coefficient.  Thus 
\begin{equation} 
\kappa_{\nu^\prime} = c(s)r_e^2 k_e (\nu_o/\nu^{\prime})(\nu_B/\nu^\prime)^{(s+2)/2} \; , 
\label{kappa} 
\end{equation} 
where $\nu_o = c/r_e$ and $r_e$ is the classical electron radius.  The function $c(s)$ is tabulated in Gould (\markcite{gould79}1979), and $k_e = (s-1)N_e \gamma_0^{s-1}/V_{\rm bw} = (s-1) r n \gamma_0^{s-1}/(3\Delta^\prime )$, using eq.(\ref{edist}). Defining the absorption frequency in this case by the value of $\nu$ satisfying $dL_\nu /d\nu = 0$, noting that $L_\nu \propto \nu^{5/2}[1-\exp(-t)]$ and $t =\kappa_{\nu^\prime}\Delta^\prime$, we obtain for the observed SSA frequency the expression 
\begin{equation} 
\nu_a ({\rm Hz}) = [{(s-1)c(s) r_e^2 r n \gamma_0^{s-1}\nu_o\over 3 t_s}]^{2/(s+4)}\;\nu_B^{(s+2)/(s+4)}\;\Gamma \;, {\rm when} \; \nu_0 < \nu_a < \nu_1\;. 
\label{nuabs1} 
\end{equation} 
The term $t_s$ is the solution to the transcendental equation $\exp(-t_s) = 5/[5+(s+4)t_s]$ derived in the slab approximation.  Representative values are $t_s = 0.35, 0.50, 0.61$ for $s = 2, 5/2,$ and 3, respectively. 
 
The synchrotron spectral luminosity $\elnumax $ at frequency $\nu_0$ is given by $\elnumax = N_e m_ec^2 \sigma_T \Gamma B/(3e)$ (\markcite{spn98}SPN98). This holds irrespective of the level of Compton losses, because $\nu_0 \elnumax$ represents the measured synchrotron power from $N_e$ electrons with Lorentz factor $\gamma\approx \gamma_0$.  Consequently, the synchrotron spectral power when $\nu_a < \nu_0$ is  
\begin{equation} 
L_\nu = \elnumax \times\, \cases{ ({\nu_a\over\nu_0})^{1/3}({\nu\over \nu_a})^2\;, & for $\nu < \nu_a$ \cr 
	 ({\nu\over \nu_0})^{1/3} & for $\nu_a < \nu < \nu_0$,\cr 
		({\nu\over \nu_0})^{-(s-1)/2} & for $\nu_0 < \nu < \nu_1$ \cr 
		({\nu_1\over \nu_0})^{-(s-1)/2}({\nu\over\nu_1})^{-p/2} & for $\nu_1 < \nu < \nu_2$ \cr	}\; , 
\label{sync1} 
\end{equation} 
and $\nu_a$ is given by eq.(\ref{nuabs}). 
When $\nu_0 < \nu_a < \nu_1$, $\nu_a$ is given by eq.(\ref{nuabs1}), and the first two cases in eq.(\ref{sync1}) should be replaced by 
\begin{equation} 
L_\nu =  
\elnumax \times\, \cases{ ({\nu_0\over \nu_a})^{5/2}({\nu_a\over \nu_0})^{-(s-1)/2}({\nu\over \nu_0})^2\;, & for $\nu < \nu_0$ \cr 
	 ({\nu\over \nu_a})^{5/2}({\nu_a\over \nu_0})^{-(s-1)/2} & for $\nu_0 < \nu < \nu_a$,\cr	}. 
\label{sync2} 
\end{equation} 
The last two cases of eq.(\ref{sync1}) also hold at frequencies $\nu >\nu_a$ when $\nu_0 < \nu_a < \nu_1$. 
 
\section{Synchrotron Self-Compton Emission} 
 
Moderski et al.\ (\markcite{msb99}1999) show that Compton processes are energetically negligible when $\e \equiv \e_e \e_{\rm rad} \ll \lambda \e_B$ (see also  Sari, Narayan, \& Piran \markcite{snp96}1996). This is a conservative estimate because it assumes that all scattering takes place in the Thomson regime. Consequently SSC processes can only be neglected with certainty if the energy transfer from protons to electrons is small and the magnetic field energy density is near equipartition. To explain the prompt gamma-ray burst emission in an external shock scenario, it is necessary to have $\e_B \ll 1$ in order to be in the weakly cooled regime, in accord with measurements of the hard X-ray/soft gamma-ray spectral indices in GRB spectra during the prompt phase (see discussion in Chiang \& Dermer \markcite{cd99}1999). SSC emission must therefore be treated in this case, and also when $\e_e \gtrsim 0.1$, which is an important regime to consider for good radiative efficiency in GRB blast waves. 
 
We employ a very simple treatment for the SSC process by approximating the synchrotron emission by a $\delta$-function distribution at the comoving frame frequency $\nu_1/\Gamma$. This is the frequency of the peak of the power output if $2< p <3$.  As noted earlier, the relative energy radiated in Compton and synchrotron photons is, neglecting Klein-Nishina effects, equal to $u_{\rm ph}/u_B \cong L_{\rm syn}/(u_B 4\pi r^2 c)$,  where $L_{\rm syn}$ is the synchrotron power in the comoving frame. The synchrotron power is 
$$L_{\rm syn} = {4\over 3}\;c\sigma_T u_B \int _{\gamma_0}^{\gamma_2} d\gamma N(\gamma)\gamma^2 = $$ 
\begin{equation} 
{4\over 3} (s-1) c\sigma_T u_B N_e \gamma_o^{s-1}\;[\gamma_1^{3-s}({1\over 3-s} - {1\over 2-p})+{\gamma_2^{2-p}\gamma_1^{p+1-s}\over 2-p} - {\gamma_0^{3-s}\over 3-s}]\;. 
\label{elsyn} 
\end{equation} 
 
Because of the strong reduction in scattering efficiency when photons are scattered in the Klein-Nishina limit, we shut off the integration when $\gamma h\nu_1/(\Gamma m_e c^2) > 1$, or when $\gamma = \bar  \gamma = \Gamma m_e c^2/(h\nu_1)$. This implies that no photons with frequencies $\nu_{\rm KN} > (\Gamma m_ec^2/h)^2 \nu_1^{-1}$ are emitted. The energy density of photons that are efficiently scattered in the Thomson regime is obtained by repeating the exercise in eq.(\ref{elsyn}), but integrating over the range $\gamma_0 < \gamma < \bar \gamma$ rather than over $\gamma_0 < \gamma < \gamma_2$.  This yields a ratio $f_{KN}$ of the Compton power which approximately takes into account Klein-Nishina effects to that without Klein-Nishina effects.  This ratio is roughly given by $f_{\rm KN} = 1$ if $\bar\gamma/\gamma_1 > 1$, and $f_{\rm KN} = (\bar\gamma/\gamma_1)^{3-s}$ if $\bar\gamma/\gamma_1 < 1$. Because $\bar\gamma$ is the electron Lorentz factor for which Klein-Nishina effects become important, it is also necessary that $\bar\gamma > \gamma_0$ for significant Compton emission. 
 
The spectral power of the Compton-scattered radiation at frequency $\nu_0^{\rm C} = \gamma_0^2\nu_1$ is  
\begin{equation} 
L_{\nu,{\rm max}}^{\rm C} = {4\over 3}\; {c \sigma_{\rm T} u_{\rm ph}\gamma_0^2\Gamma^2 \over \nu_0^{\rm C}}  N_e = {u_{\rm ph}\over u_B}{\nu_0\over \nu_0^{\rm C}} \elnumax = \gamma_1^{-2}\; {u_{\rm ph}\over u_B}\; \elnumax . 
\label{elnumaxc} 
\end{equation} 
The SSC component in this approximation is therefore given by 
\begin{equation} 
L_\nu^{\rm C} = \elnumax^{\rm C} \times\, \cases{ ({\nu\over\nu_0^{\rm C}})^{1/3}, & for $\nu < \nu_0^{\rm C}$ \cr 
		({\nu\over \nu_0^{\rm C}})^{-(s-1)/2} & for $\nu_0^{\rm C} < \nu < \nu_1^{\rm C} = \gamma_1^2 \nu_1$ \cr 
		({\gamma_0\over \gamma_1})^{(s-1)}({\nu\over\nu_1^{\rm C}})^{-p/2} & for $\nu_1^{\rm C} < \nu < \nu_2^{\rm C} = \gamma_2^2 \nu_1$ \cr	}\; , 
\label{comp1} 
\end{equation} 
provided $\nu_{\rm KN} > \nu_2^{\rm C}$.  When this is not the case, the SSC emission is set equal to zero at $\nu > \nu_{\rm KN}$. 
 
\section{Results} 
 
Figure 1 compares spectral energy distributions (SEDs) obtained from the semi-analytic formulas derived above with those calculated using a detailed numerical simulation of GRB blast waves.  The numerical model is described by Chiang \& Dermer (\markcite{cd99}1999) and Dermer et al.\ (\markcite{dcm99}1999).  The parameters in this calculation represent a standard set chosen to produce a bright, prompt GRB with a peak gamma-ray luminosity $\sim 10^{51}$ ergs s$^{-1}$, a duration $\sim 10$ s, a $\nu F_\nu$ peak synchrotron frequency $\nu_{\rm pk}$ corresponding to photon energies of several hundred keV, and a hard spectrum approaching $\alpha = -1/3$ at $\nu \ll \nu_{\rm pk}$, where the flux density $F(\nu)\propto \nu^{-\alpha}$. For this calculation, a fireball with total energy $E_0 = 10^{54}$ ergs expands to produce an uncollimated blast wave (effects of  blast-wave collimation are numerically treated by Moderski et al.\ \markcite{msb99}1999, Panaitescu \& M\'esz\'aros \markcite{pm99}1999, and  Dermer et al.\ \markcite{dcm99}1999).  For the reasons given above, a very small magnetic-field parameter of $\epsilon_B = 10^{-4}$ is used. The injection index $p= 2.5$ is used to match spectral indices inferred from afterglow spectra at optical and X-ray frequencies.  The curves are labeled by the base 10 logarithm of the observer time, and the numerical results are highlighted by data points. 
 
As can be seen from Fig.\ 1, the behavior of the blast wave emissions at soft gamma-ray energies and below is qualitatively reproduced by the analytic expressions.  The synchrotron emission reaches a peak at early times and at high photon energies.  The deceleration of the blast wave causes the observed synchrotron fluxes and $\nu_{\rm pk}$ values to decrease with time.  The fit is best during the afterglow phase at $10^4~{\rm s}\lesssim t \lesssim 10^{7}~{\rm s}$ at  frequencies $10^{14}~{\rm Hz} \lesssim \nu \lesssim 10^{18}~{\rm Hz} $.  The characterization of the high-energy gamma-ray SSC component is very poor, especially during the prompt phase.  This is due to the oversimplified way of dealing with the Klein-Nishina reduction in the cross section. Progress in improving the analytic representation of the SSC emission comes at the expense of cumbersome formulas and which furthermore still omit the effects of photon-photon attenuation that can be important at TeV energies.  Thus we do not go beyond the simple analytic treatment of the SSC component given in the previous section. 
 
The analytic SEDs also provide a poor representation of the hard X-ray spectra shortly after the prompt emission phase at $t \sim 10$-$10^3$ s.  This discrepancy has at least two causes. The first is the neglect of adiabatic losses on the electrons. These losses degrade the low-energy portion of the electron distribution function and produce a smoother low-energy cutoff in this distribution function than assumed in the analytic representation. The second is the analytic approximation for the elementary synchrotron emissivity spectrum, which consists of two connected power laws. This shortcoming is illustrated in Fig.\ 2, where we plot synchrotron emissivity spectra using the formulas of Crusius \& Schlickeiser (\markcite{cs86}1986), which have been averaged over pitch-angle and magnetic field direction, for power-law electron distribution functions with index $s$. These results are compared with an analytic representation of the synchrotron emisivity spectrum derived in a $\delta$-function approximation (Dermer, Sturner, \& Schlickeiser \markcite{dss97}1997).  The $\delta$-function approximation yields the expression of \markcite{spn98}SPN98 for the synchrotron emission except for an overall normalization of a factor $2/(p-1)$. Fig.\ 2 shows that the analytic results overestimate the flux by as much as a factor of $\sim 4$ at the low-energy break where the analytic power-law connects to a power law with energy index $\alpha = -1/3$ extending to lower frequencies. At the high-energy endpoint of the synchrotron spectra, Fig.\ 2 shows that even larger discrepancies between the numerical and analytic results are obtained, though these are usually concealed by the SSC component. 
 
The effects of adiabatic losses on the low-energy distribution function becomes less important at later times. Thus the flux discrepancies in the low-energy portion of the synchrotron spectra decrease with time, but they still can differ by an order-of-magnitude.  
 
Fig.\ 3 shows the effect of considering a stronger blast-wave magnetic field in the model.  All parameters are the same as in Fig.\ 1, except that $\epsilon_B = 0.1$. With this stronger field, the GRB would produce a brighter observed synchrotron  emission spectrum, though most of its power would be carried by photons with higher energies.  In this case, the bulk of the synchrotron power is emitted in the form of photons with energies of $\sim 50$ MeV.  Note, however, that this model GRB would be dimmer in the BATSE triggering range than the model GRB shown in Fig.\ 1.  The strength of the SSC component is much weaker than in the model of Fig.\ 1, particularly in the afterglow phase. Although the overall behavior of the numerical simulation results are reproduced by the analytic model, large discrepancies still persist for the same reasons as previously discussed.  In fact, the analytic model at optical and X-ray energies with SSC effects included is not much better than a model which neglects them, as indicated by the $10^5$, $10^6$ and $10^7$ s curves. 
 
The analytic representation of the SEDs improves when smaller values of $\epsilon_e$ are used.  Fig.\ 4 shows the effects of reducing $\epsilon_e$ to a value of 0.1, with all other parameters the same as before.  This has the effect of further reducing the relative importance of the SSC component.  Nevertheless, order-of-magnitude discrepancies between the analytic and numerical models remain at the endpoints of the spectra.  
 
Figs.\ 5a and 5b compare analytic and numerical SEDs for a model GRB with parameters derived by Wijers \& Galama (\markcite{wg99}1999) to fit afterglow data from GRB 970508. Here the total energy injected into the fireball is $E= 3.5\times 10^{52}$ ergs, which is considerably smaller than the value used in the previous cases, and the density of the external medium is very dilute with $n = 0.03$ cm$^{-3}$.  Consequently the SSC component becomes even less important than before. The analytic expressions provide a reasonable representation of the numerical model in the optical/X-ray regime, though the discrepancies at radio frequencies remain significant, as is clear from Fig.\ 5b. This model also does not explain the flux of GRB 970508 in its prompt phase, which is much brighter than calculated here.   
 
The light curves from the semi-analytic and the numerical models for the Fig.\ 5 parameters are shown in Fig.\ 6 in an $L_\nu$ representation.  For this example, there is good agreement between the optical light curves. The temporal breaks due to the peak of the synchrotron emission spectrm from the lowest energy electrons passing into the specified waveband are much smoother than the analytic model, and discrepant in flux by an order-of-magnitude.  The cooling break, which is clearly defined in the V band light curve at $t \approx 10^7$ s in the analytic model, is almost indistinct in the numerical calculation. 
 
\section{Discussion} 
 
We have examined the accuracy of analytic formulas used to model GRB afterglow radiation by comparing results from these expressions with calculations obtained with a numerical simulation model that treats the most important processes in relativistic blast waves. The advantage of the analytic model is that it can be easily used to model data and extract parameter values, but its accuracy is uncertain.  By contrast, the numerical model contains all relevant processes and should therefore be used to fit multiwavelength GRB data, but the available numerical model (Chiang \& Dermer \markcite{cd99}1999) requires $\sim $ an hour (for $\epsilon_B \ll 1$) to a day (for $\e_B\sim 0.1$-1) to run on a SparcStation. This is too long to explore parameter space as is required when fitting data.  For fits to data, it seems most useful to establish general parameters using the analytic model, and then adjust parameters in the numerical simulation to obtain a precise fit to data.  
 
We find that the analytic results are reasonably accurate (within a factor of $\sim 3$) for modeling X-ray/optical data in the afterglow phase (i.e., several hours to many days after the GRB event), but is inaccurate in the prompt and early afterglow period, especially when $\epsilon_e \gtrsim 0.1$ and $\epsilon_e/\epsilon_B \gg 1$.  Endpoint effects in the IR/radio and gamma-ray regimes, and the simplified treatment of the SSC process which is important at gamma-ray energies, limit the accuracy of the analytic model to no better than an order-of-magnitude.  As expected, the spectral and temporal breaks in the analytic approxmations are much sharper than found in the numerical calculations.  The use of the analytic model to provide simple fits to optical and X-ray afterglow data is generally warranted, but the analytic model should be applied with caution to radio data or optical and X-ray data in the prompt and early afterglow phases. 
 
\acknowledgments{We thank Professor Peter M\'esz\'aros for comments on the manuscript. This work is supported through the NASA Astrophysical Theory Program (DPR S-13756G) and the Office of Naval Research. The work of M.B. is supported by NASA through Chandra Postdoctoral Fellowship grant number  PF9-10007 awarded by the Chandra X-ray Center, which is operated by the Smithsonian Astrophysical Observatory for NASA under contract NAS8-39073.}

\eject 
 
\setcounter{figure}{0} 
 
\begin{figure} 
\centerline{\epsfxsize=0.8\textwidth\epsfbox{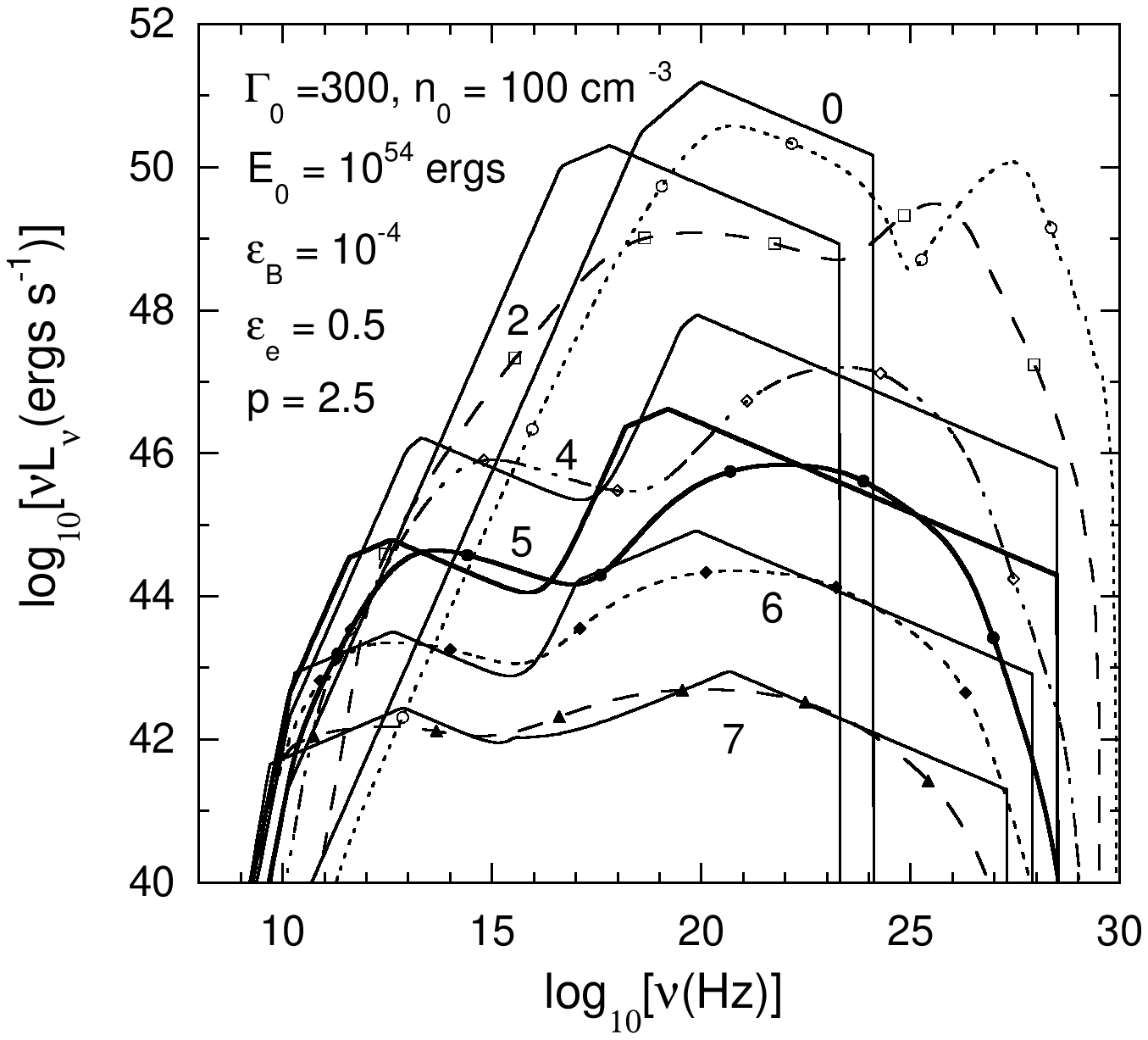}} 
\caption[]{Calculations of $\nu L_\nu$ spectral energy distributions emitted by a relativistic blast wave with initial Lorentz factor $\Gamma_0 =300$.  Curves are labeled by the base 10 logarithm of the observing time in seconds.  The analytic and numerical models are shown, with the latter curves identified by data points. The other parameters of the calculation are shown in the legend.} 
\end{figure} 
 
\begin{figure} 
\centerline{\epsfxsize=0.8\textwidth\epsfbox{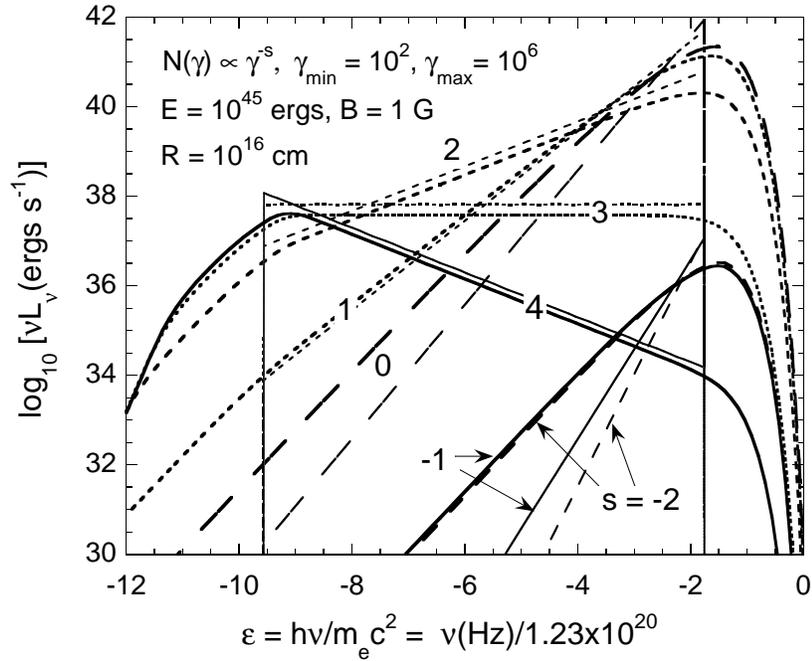}} 
\caption[]{Comparison between a $\delta$-function approximation to the synchrotron emissivity spectrum for a spherical emitting region of radius $R=10^{16}$ cm, and  the synchrotron emission obtained by averaging over pitch angles and magnetic field directions.  Curves are labeled by $s$, the index of the steady-state number distribution of electrons, which are assumed to be distributed as a power law between $\gamma_{\rm min} = 100$ and $\gamma_{\rm max} = 10^6$. The total energy of the electrons is $E=10^{50}$ ergs, and the mean magnetic field strength is $B= 1$ G. The low-energy power-law branch of the analytic approximation is not shown.} 
\end{figure} 
 
\begin{figure} 
\centerline{\epsfxsize=0.8\textwidth\epsfbox{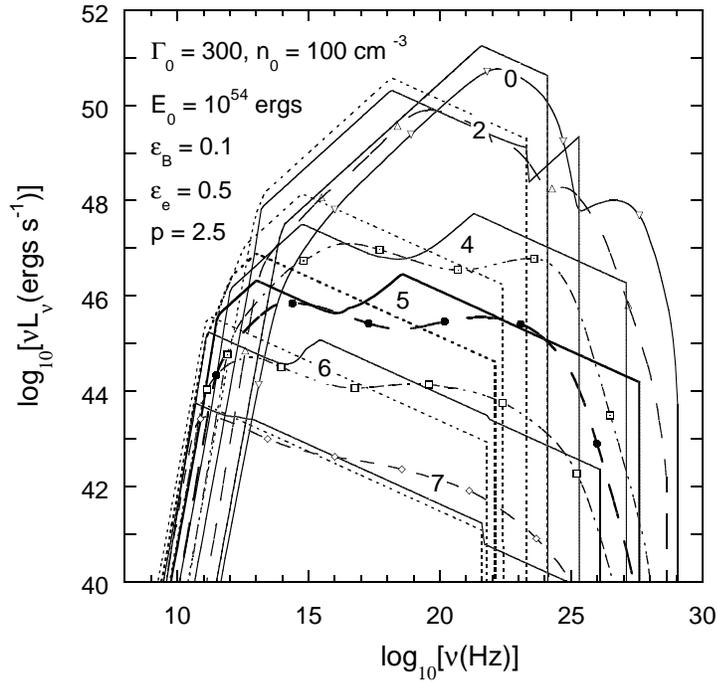}} 
\caption[]{Same as Fig.\ 1, but with the magnetic equipartition parameter $\epsilon_B = 0.1$. 
Two analytic models, one with SSC (solid curves) and one without SSC (dotted curves) processes included, are shown. The curves are identical for the  t $= 1$ s case. } 
\end{figure} 
 
\begin{figure} 
\centerline{\epsfxsize=0.8\textwidth\epsfbox{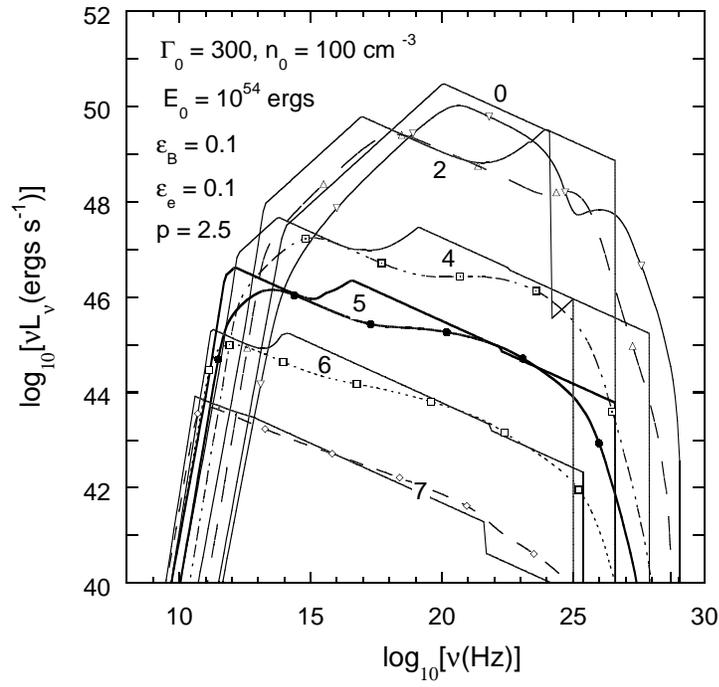}} 
\caption[]{Same as Fig.\ 1, but with $\epsilon_B = 0.1$ and $\epsilon_e = 0.1$.} 
\end{figure} 
 
\begin{figure} 
\centerline{\epsfxsize=0.5\textwidth\epsfbox{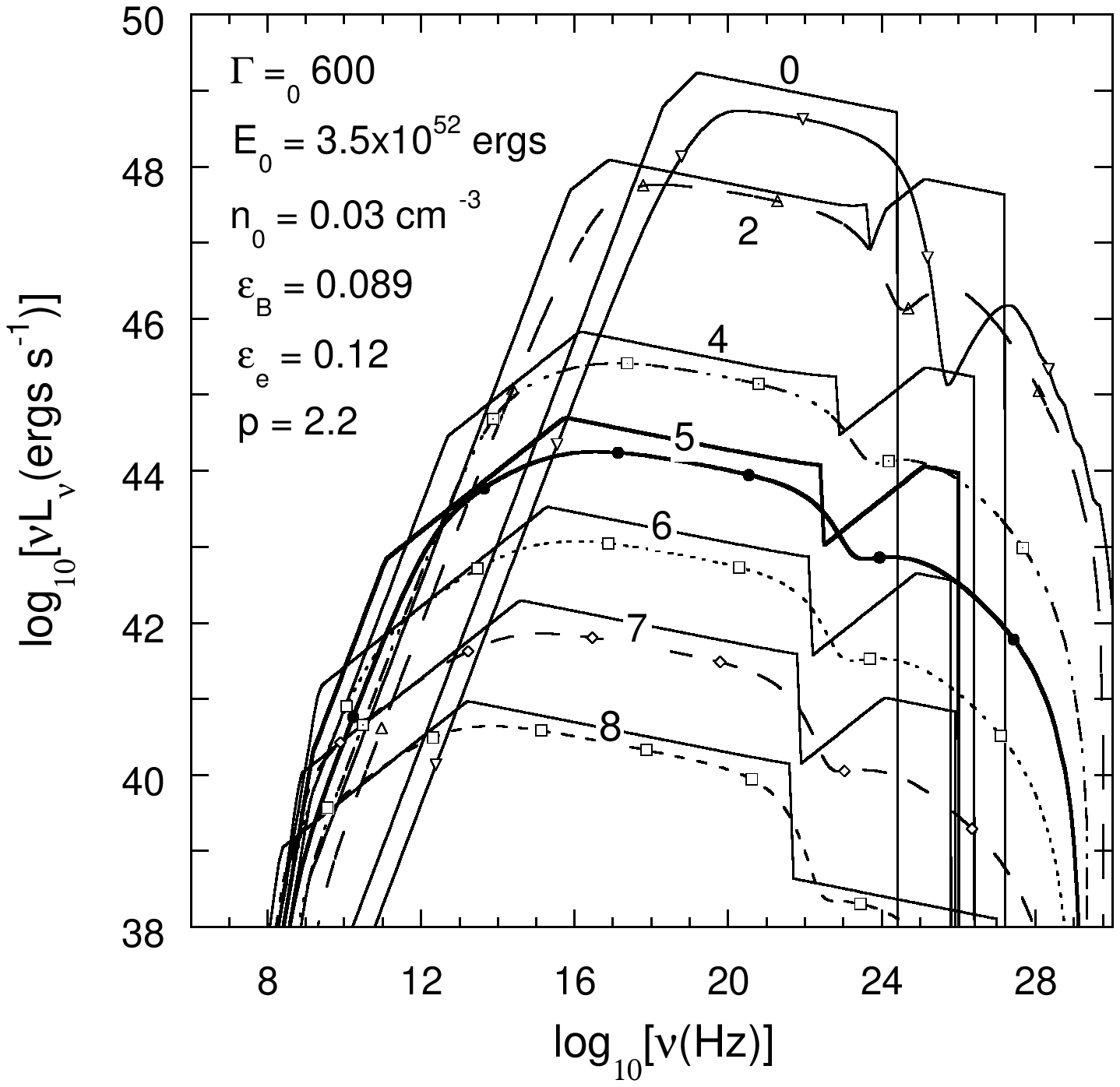} 
            \epsfxsize=0.5\textwidth\epsfbox{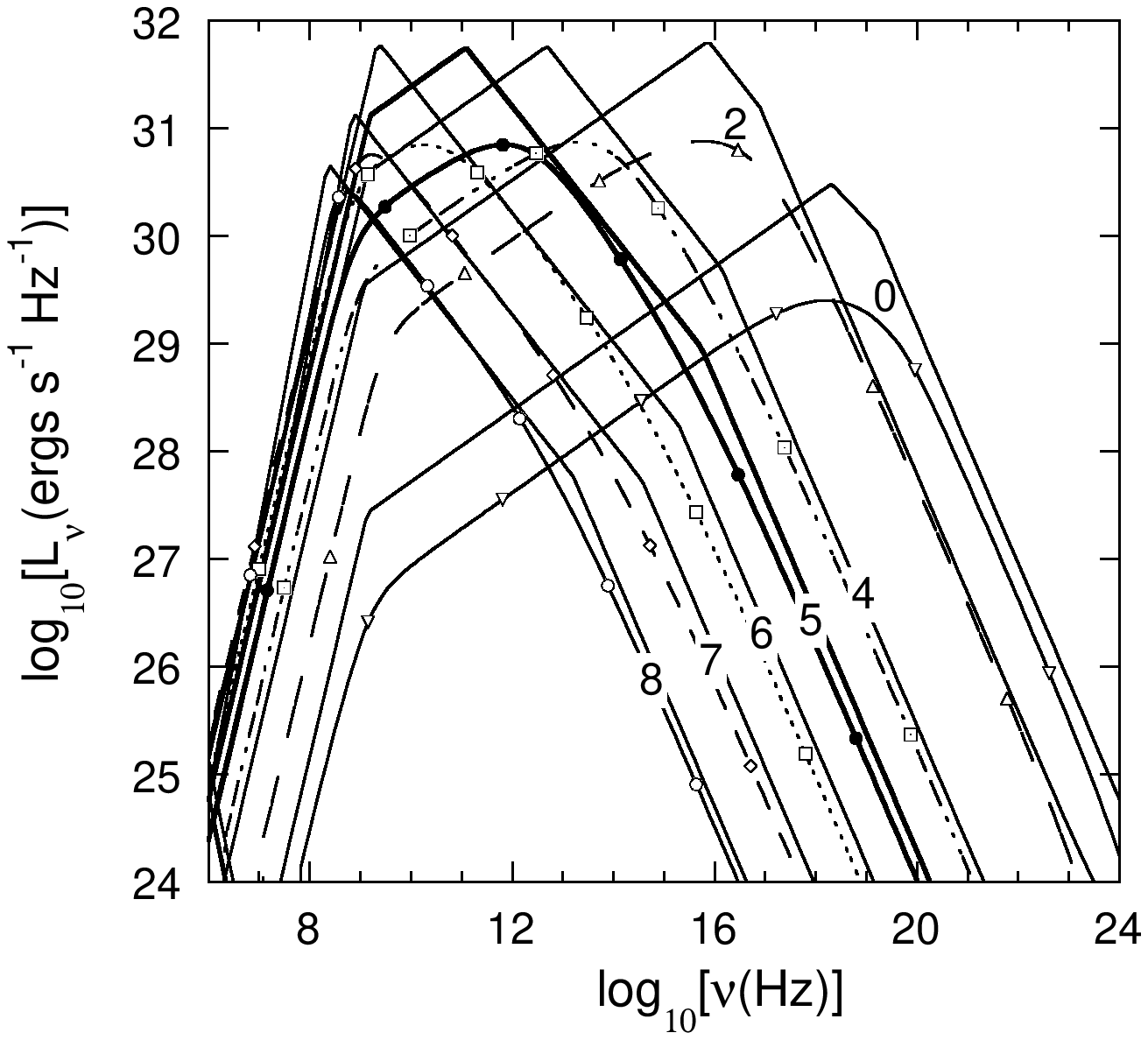}} 
\caption[] {Same as Fig.\ 1, but with parameters taken from the fit to GRB 970508 by Wijers \& Galama \markcite{wg99}(1999). (a) Comparison of SEDs in a $\nu L_\nu$ representation.  (b) Comparison of SEDs in an $L_\nu$ representation.} 
\end{figure} 
 
\begin{figure} 
\centerline{\epsfxsize=0.8\textwidth\epsfbox{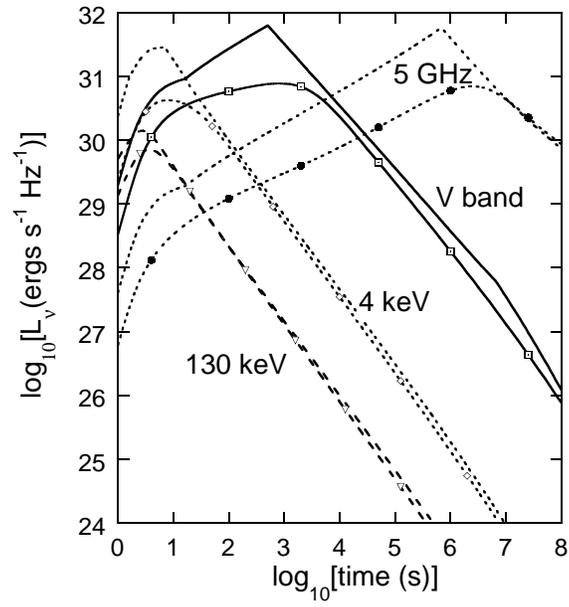}} 
\caption[]{Analytic and numerical light curves calculated at radio, optical, X-ray and soft gamma-ray energies for the case shown in Fig.\ 5. } 
\end{figure} 
 
\end{document}